# VOYAGERS 1 AND 2 IN A SHRUNKEN AND SQUASHED HELIOSPHERE

**W.R. Webber[1] and D.S. Intriligator[2]**

1. New Mexico State University, Department of Astronomy, Las Cruces, NM 88003, USA
2. Carmel Research Center, Space Plasma Laboratory, Santa Monica, CA 90406, USA

**ABSTRACT**

16          We have extended our earlier calculations of the distance to the Heliospheric

17     Termination Shock (HTS) - which covered the period from the launch of V1 and V2 in

18     1977 to 2005 - to the period from 2006 to 2011.  During this latter period the solar wind

19     speed, ram pressure and magnetic field decreased to the lowest levels in recent history,

20     related to the sunspot minimum in 2008-2009.  The HTS distance has decreased

21     correspondingly so that V1, which was crossed by the HTS at 94 AU in late 2004, would

22     now, in early 2011, be expected to reach the HTS at a distance ~80 AU, when the HTS

23     distance would be expected to be at its minimum.  Similarly V2, which was crossed by

24     the HTS at 84 AU in mid 2007, would, in early 2011, reach the HTS at a distance of only

25     74 AU.  These distances, in early 2011, are ~15% less than those at which V1 and V2

26     initially reached the HTS.  The distance to the Heliopause (HP) is more uncertain but

27     recent calculations place its equilibrium distance at between 1.4-1.6 times the HTS

28     distance.  Allowing for an additional 1 year for the HP to reach its equilibrium minimum

29     distance relative to the HTS would mean that, assuming this distance remains a constant

30     fraction larger than the HTS distance, the HP distance would be at its minimum distance

31     of (1.4-1.6) x 80 AU = 112-128 AU at V1 in early 2012.  At this time V1 will be at a

32     distance of ~120 AU so that there is a possibility that V1 could cross the HP and enter

33     interstellar space at the time $2012.0 \pm 1$ year.  If the crossing does not happen during this

34     time period, then it is unlikely that V1 will reach this defining boundary before about

35     2016 because of the expected outward motion of the HTS and the HP towards their more

36     normal distances of 85-96 and ~120 -140 AU coincident with the maximum of the new

37     sunspot cycle.

38

## Introduction

The location of HTS is determined by the pressure balance between the outward flowing solar wind and the plasma and magnetic field properties of the local interstellar medium. V1 first reached the HTS in late 2004 at a distance of 94 AU. This provided a normalization point for determining the HTS distance with time that was utilized in a simple empirical model to estimate the effect of solar wind pressure variations on the shock location using the available solar wind plasma data from V2 (Webber, 2005). Subsequently more sophisticated models were employed to trace the HTS location (e.g., Richardson, et al., 2006a) and to also examine the heliosheath region beyond the HTS (Washimi, et al., 2007, 2010). Basically the prediction of all of these models is similar with regard to the variations of the HTS distance with time. Over a comparable time period from about 2002 to 2007, covering a period from maximum to minimum HTS distance, the maximum variation in each model was ~10-13 AU.

At the location of V1 the HTS distance was estimated to be ~85 AU in 2001 (just beyond or outside of the V1 distance at that time) and moved rapidly outward, still just outside of the V1 location, between 2002-2004 (Webber, 2005). Then near the end of 2004 the HTS started an inward movement and the HTS crossed V1 at 94 AU. The HTS continued to move inward rapidly, and reached a distance of ~87 AU by the end of 2005, so that V1 had moved ~10 AU beyond the shock in only 1 years time (Webber, et al., 2007) (the average speed of V1 is 3.6 AU/year)!

The object of this paper is to determine the distance of the HTS at both V1 and V2 during the period just before V2 reached the HTS in 2007, again using V2 plasma data (see also Washimi, et al., 2010) and then in the time period from 2007-2011, to determine the HTS distance at V1, but now using the plasma data at 1 AU from the SWEPAM instrument on ACE (http://www.srl.caltech.edu/ACE/) and the OMNI plasma data (http://omniweb.gsfc.nasa.gov/ form/dx1.html) since the V2 plasma data are no longer useful in this regard after the HTS crossing. This later time period is one in which the cosmic ray intensities reached an unpredicently high maximum and the solar wind speed and pressure and the interplanetary magnetic field reached unusually low values corresponding to the sunspot minimum in 2008-2009 (e.g., McDonald, Webber and

69   Reames, 2010; Mewaldt, et al., 2010).  For this later time period we shall also estimate

70   the distance to the heliopause (HP), assumed to be at a distance beyond the HTS that

71   remains a constant fraction of the HTS distance from the Sun (e.g., Mueller, et al., 2006).

72   **Data**

73          From the beginning of 2005 to the time the HTS crossed V2 we use the daily

74   average density and solar wind speed data from the V2 MIT plasma experiment to

75   calculate   the   time   history   of   the   solar   wind   ram   pressure,   $nV^2$   (see

76   http://web.mit.edu/afs//athena/org/s/space/www/                    voyager/voyager_data/

77   voyager_data.html).  The 5 day running average pressure is corrected to a constant 1 AU

78   using an $r^{-2}$ dependence for the density (no correction is made for a possible radial

79   dependence of V).  This pressure is shown in Figure 1 for the time period from 2002-

80   2007 when V2 was between 75-85 AU.  The main features of these data are the large

81   pressure wave seen at 2006.17 and two decreases in pressure seen about 0.26 and 0.16

82   years prior to the HTS crossing of V2.  The blue and red lines in the figure show

83   extrapolations of the pressure that may have been seen at V2 during the latter half of

84   2007 if the HTS were not present.  They will be discussed later.

85          After about 2007.4, the V2 data cannot be reliably used to indicate the solar wind

86   pressure inside the HTS.  For these time periods we use the SWEPAM data from the

87   ACE spacecraft at the Earth.  The day to day variations of this pressure, caused mainly by

88   density variations, is large so in Figure 2 we show the 27 day average solar wind ram

89   pressure at the Earth additionally smoothed with a 5 period running average.  These data

90   show several broad time periods of increased pressure during the 11-year solar activity

91   cycle and in particular after 2008.0, show a rapid decrease in the overall solar wind

92   pressure as the solar activity approaches its minimum.  By the end of 2009 this ram

93   pressure is less than half of the average pressure over the earlier part of the 11-year solar

94   cycle.  This time period is consistent with the low average values of the interplanetary

95   magnetic field and the unusually low amount of solar modulation of galactic cosmic rays

96   as described by McDonald, Webber and Reames, 2010 and Mewaldt, et al., 2010.

97          In Figure 3 we show the solar wind ram pressure measured at the V2 spacecraft

98   over the same time period and with the same smoothing.  The Earth based SWEPAM

99      data in this figure are delayed by 0.87 years to correspond to an average solar wind

100     transient time between the Earth and V2. The broad temporal features seen at the Earth

101     line up well with those seen at V2 using this delay.

102          In Figure 4 we show the ratio of the ram pressure measured at V2 (corrected to 1

103     AU) and at the Earth, delayed by 0.87 years, over the entire time period of 1998-2007,

104     nearly a complete 11-year solar cycle. This ratio has an average value slightly less than

105     1 over this time period, as might be expected from a slowing down of the solar wind

106     speed and also shows the large temporal variations exhibited in Figure 3. The ratio of 1,

107     as shown by the dashed line in Figure 4, is used to approximate the temporal variations in

108     the solar wind ram pressure that would be seen near the HTS after 2008.0 from the

109     smoothed variations observed at the Earth by SWEPAM. These Earth data are extended

110     to late 2010 by using the solar wind data from OMNI.

111     **Responses of the HTS to the Solar Wind Pressure Variations**

112          The fluctuating solar wind pressure variations make it difficult to calculate

113     theoretically the expected HTS distance with time at any one location. Certainly, on a

114     longer term time scale the HTS will be located near the point where there is equilibrium

115     between the outward solar wind pressure and the inward local interstellar plasma and

116     magnetic field pressure. On a shorter time scale the HTS location will also move inward

117     and outward corresponding to the short term pressure variations in the solar wind.

118     Complex latitude and longitude asymmetries in the solar wind pressure will result in a

119     non uniform HTS distance with respect to both heliospheric latitude and longitude.

120          For our calculation we consider both the short term (~26 day) and long term (~1

121     year or longer) pressure variations to calculate a typical HTS distance near the solar apex

122     within ± 30 degrees of the heliospheric equator. This includes the locations of V1 and

123     V2. Possible latitude effects of the solar wind ram pressure and therefore the HTS

124     distance, which could be important, are not considered in this analysis.

125          The details of the calculation of the HTS location with time are discussed in

126     Webber, 2005. Briefly in summary here we note that this calculation considers both long

127     term and short term pressure variations as well as the inward and outward speeds of the

128     HTS in each 26 day interval in response to the solar wind pressure variations. For the

129     long-term HTS variations we take a 3 period moving average of the 6-month average

130     solar wind pressure data.  This time sequencing is similar to that used by Izmodenov, et

131     al., 2003, in their calculation of the HTS distance in which they use a 5 period moving

132     average of the 6 month average solar wind data to calculate the HTS location and its

133     changes with time.  For the inward and outward speed of the HTS in response to these

134     longer term pressure variation we used as a guide the work of Wang and Belcher, 1999.

135     The shorter term 26 day pressure variations are superimposed on the longer term

136     variations.  Here we recognize that the average pressure difference between successive 26

137     day periods may change by a factor of ~2 or more.  These pressure differences are usually

138     related to the passage of large interplanetary shocks (e.g., GMIRS).  For the inward and

139     outward speeds during these time periods, we make use of calculations by Whang and

140     Burlaga, 1993 and Lu, et al., 1999 for interplanetary shocks interacting with the HTS.  In

141     some cases, when the pressure difference between two successive 26 day intervals is

142     large, outward and inward speeds of the HTS of at least 100 km·sec$^{-1}$ are estimated during

143     a 26 day period.

144     **Variations of the HTS Distance at V1 and V2 during the 2002-2007 Time Period**

145     In Figure 5 we show the calculated distance of the HTS at the V1 location for the

146     2002-2007 time period using our method.  Also shown in Figure 5 are the radial locations

147     of V1 and V2 as a function of time.  For V1 we observe that throughout 2002, 2003 and

148     2004 this spacecraft is just beyond of the estimated HTS location.  This time period and

149     its relationship to the large intensities of MeV protons seen at V1 has been discussed in

150     Webber, 2005 (see also Washimi, et al., 2007).  At the end of 2004, the HTS crossed V1

151     during a period of rapid inward movement after the so-called Halloween 2003 shocks

152     (October-November, 2003 at the Earth) reached the HTS.

153     The times labeled 1A, 1B and 2 in Figure 5 show the effects of the arrival of

154     specific large shocks at the HTS and its subsequent outward (inward) movement.  The net

155     effect of these interplanetary shocks was to temporarily move the HTS outward by as

156     much as 5 AU over 2 or 3 - 26 day time periods.  Shock 1B is the Halloween shock

157     (Intriligator, et al., 2005), shock 2 is the large shock seen at V2 in March 2006

158     (Richardson, et al., 2006b).

159     The behavior of the HTS at the V2 location is also shown in Figure 5.  This curve,

160     which essentially matches the V1 curve, is normalized so that the HTS crossed V2 at the

161     observed time of 2007.66.  This normalization factor, taken here to be 0.92, is a measure

162     of the asymmetry of the heliosphere between the N-S distances at which V1 and V2

163     reached the HTS (see Stone, et al., 2008; Opher, 2010; Washimi, et al., 2010).  This

164     asymmetry factor is of importance for comparing V1 and V2 cosmic ray intensity

165     observations to determine a radial gradient, for example, and is often neglected.  These

166     two spacecraft are separated by about 20 AU in radial distance over this time period and

167     this "squashing" of the heliosphere by ~8-10% or 8-10 AU near the location of the HTS

168     significantly changes the values of the radial gradient determined.

169     Notice in Figure 5 that after about 2007.5 the HTS location at both V1 and V2 is

170     rapidly moving inward.  This inward motion arises from the fact that the solar wind

171     pressure measured at V2 suddenly begins to decrease at about 2007.4, well before the

172     arrival of V2 at the HTS itself.  A second further decrease in solar wind pressure occurred

173     at about 2007.5 so that in the time period just prior to the arrival of V2 at the HTS, the

174     solar wind ram pressure became less than half of the average pressure observed at V2

175     throughout 2006 and 2007 (see Figure 1 and Richardson, et al.,2008).

176     The HTS distance at V2 shown in Figure 5 is based on the V2 solar wind plasma

177     data up to the time of arrival of V2 at the HTS at 2007.66, the same as the Washimi, et

178     al., 2010, calculation.  If we believe that these solar wind data are affected by structures

179     in front of (sunward of) the HTS after about 2007.4, then we have to make assumptions

180     about the solar wind pressure at V2 for the remainder of 2007.  Possible assumptions are

181     shown by the red curve and blue curves a and b in Figure 1.  In each case the solar wind

182     pressure is taken to be that measured by V2 up to 2007.4.  For the red curve the shock

183     from the December, 2006 events at the Earth is taken to arrive at V2 at 2007.50, the time

184     of a large magnetic field increase indicative of a GMIR (Burlaga, et al., 2008) as well as a

185     large decrease of >70 MeV cosmic rays (Webber, et al., 2009).

186     This December, 2006 event has also been propagated through the heliosphere by

187     Intriligator, et al, 2010.  Their calculated arrival time of this shock is about 2007.72 (day

188     262) and the convective plasma speed increase in the heliosheath at this time is 1.15

189    times (from 325 km·s⁻¹ to 375 km·s⁻¹). The HAFSS model used (as discussed in

190    Intriligator, et al., 2010a) does not take into account the fact that the December, 2006 IP

191    shock is propagating beyond the HTS. HAFSS assumes the IP shock is always in the

192    solar wind. Also including the effects of slowing down, we estimate the interplanetary

193    shock as calculated by Intriligator, et al., 2010a, arrived at V2 in the 26 day time interval

194    centered on 2007.86 ± one 26 day interval, with a shock pressure increase of 1.3 times.

195    An enlarged profile of the HTS behavior in the year 2007 for these different

196    possibilities is shown in Figure 6. Instead of a rapidly decreasing HTS distance at the

197    time of the V2 HTS crossing, we now find that the HTS distance is almost constant

198    throughout the period before and after the shock crossing for all three possibilities. Thus

199    the HTS first crossed V2 at 2007.66 and V2 is slightly beyond the HTS just before

200    2007.86. When the interplanetary shock predicted by Intriligator, et al., 2010a, reached

201    the HTS, just inside V2, it pushed the HTS out to near the location of V2 during the

202    2007.86 - 26 day interval. Thus a second HTS encounter or near encounter could occur

203    at that time. The HTS moved out ~1 AU at this time in accordance with the assumed

204    solar wind pressure increase of ~1.3 times.

205    We note that Intriligator, et al., 2010a, have observed high energy solar wind ions

206    at roughly twice the average solar wind speed at V2 two times in 2007. The first time

207    was on days 242-243 in conjunction with the HTS crossing (Intriligator, et al., 2010b),

208    the 2$^{nd}$ time was between days 333-348 (2007.91-2007.95). Large transient energetic

209    plasma, TSP and ACR intensity changes were seen near these times in later 2007 by the

210    LECP and CRS instruments on V2 (Decker, et al., 2009; Intriligator, et al., 2010a). Large

211    magnetic field enhancements were also seen near the times of the particle intensity

212    changes (Burlaga, et al., 2008). Further studies are underway to verify whether these

213    times could be times of HTS crossings.

214    Subsequently, in 2008 the HTS began to move inward as the decreasing solar

215    wind pressure associated with the 11-year solar cycle propagated out to the location of

216    V2.

217    **The Distance of the HTS and HP from 2008-2012**

218    The behavior of the HTS at V1 in the time period from 2008-2012, based on the
219    SWEPAM and OMNI data at the Earth, is shown in Figure 7. The HTS moved inward,
220    slowly at first during 2008, but in 2009 and 2010 moved inward by ~4-5 AU/year,
221    actually greater than the outward speed of V1. By 2011.0 the HTS reached a location at
222    ~80 AU while at the same time V1 was at ~115.7 AU. After 2011.0 the HTS behavior
223    will depend on the solar wind pressure at the Earth after about ~2010.0. OMNI data
224    show that this pressure increases in the 1st half of 2010, and then levels off through the
225    end of 2010. By the end of 2011 the HTS distance at V1 has increased from a minimum
226    of ~80 AU to ~82.5 AU.

227    The changes in the distance to the HP during this time period are more
228    problematical. The studies of Mueller, et al., 2006, have shown that, for a wide range of
229    interstellar parameters and a wide range of HTS distances, the ratio of the HP distance to
230    the HTS distance remains roughly constant at a value ~1.4. From this study one might
231    conclude that, as the HTS moves inward during the 2008-2011 time period due to solar
232    wind pressure changes, the HP will move inward as well and that, once equilibrium is
233    established, the HP will be at a closer distance which is still ~1.4 times the new HTS
234    distance. In this scenario, then, assuming that equilibrium is established in ~1 year; (the
235    travel time of an outward moving disturbance to propagate through the heliosheath, e.g.,
236    Washimi, et al., 2010) by about 2012.0 the HP distance would be at its minimum = 1.4
237    times the HTS minimum of 80 AU, or ~112 AU and by about 2013.0, the last point at
238    which we have data, the HP would be at ~115 AU and moving outward at about the same
239    speed as V1. The curve corresponding to this minimum HP distance of 1.4 times the
240    HTS distance and a ~equilibrium time of 1 year is shown in Figure 7. The V1 trajectory
241    first intersects this curve early in 2011. There is evidence in the changing pattern of the
242    solar wind flow as of about 2010.5 that V1 may be approaching the HP (Decker, et al.,
243    2010).

244    But there is the question, how will we know when V1 is approaching or has
245    crossed the HP? On V1, without a working plasma detector, this might be more difficult.
246    From the energetic particle point of view, the intensities of ~100 KeV plasma particles
247    measured by the LECP experiment (Decker, et al., 2009), the MeV termination shock

248     particles measured by both the LECP and CRS experiments and the ACR particles
249     measured in most detail by the CRS experiment (Cummings, et al., 2010), provide
250     interesting possibilities for sensing the proximity of the HP or some equivalent "outer
251     boundary". The TSP and ACR populations have extremely high intensities in the
252     heliosheath that are characteristic of "trapped" populations. These high intensities first
253     appeared, more or less coincident with the HTS crossing and have continued, even
254     increasing intensity in the case of the ACR, up to the present time when V1 is
255     approaching 115 AU.

256     At some distance, near the outer trapping boundary, these intensities should begin
257     to decrease, with the higher energies most likely decreasing 1$^{st}$ because of their longer
258     diffusion length. Calculations relating to this problem have been made by Ferreira,
259     Potgieter and Scherer, 2006, in their model for the acceleration and propagation of ACR.
260     These calculations, above a few MeV, show that the peak intensities of ACR will occur
261     about ~5-10 AU in front of an "outer boundary". This boundary need not be the HP but
262     could be a specific "region", (e.g., the magnetic wall identified in the calculations of
263     Washimi, et al., 2010), in the structure of the heliosheath. The magnetic wall itself
264     appears to be ~10 AU inside the HP according to Washimi, et al., 2010. At the present
265     time (early 2011) the intensities of ACR above a few MeV certainly appear to have
266     reached a "peak" intensity (Cummings, et al., 2010).

267     Returning to the question of the distance of the HP during the 2008-2013 time
268     period, we also observe that there are a number of different estimates regarding the
269     distance to the HP. Relative to the HTS most of these estimates place the equilibrium HP
270     distance (in the apex direction) at between 1.4-1.6 times the HTS distance. Thus the
271     Mueller, et al., 2006, value is at the low end of this range. The work of Florinski and
272     Pogorelev, 2009, uses a HP distance of between 1.5-1.6 times the HTS distance in order
273     to describe the cosmic ray modulation in the outer heliosphere. A similar value for this
274     HP/HTS distance ratio is obtained from the studies of Washimi, et al., 2010, on the
275     characteristics and structure of the heliosheath.

276     We thus take a reasonable upper limit for the equilibrium value of the HP/HTS
277     distance ratio to be 1.6. This is shown as the upper limit of the shaded region in Figure 7.

278    V1 will not reach this limiting distance during the 2008-2012 time period.  In fact, V1

279    which is at ~121 AU near the middle of 2012, will be at a distance which is = 1.50 times

280    the HTS distance.  After early 2012 the HP distance will have already started to increase

281    in response to the increasing solar activity and solar wind ram pressure as observed by

282    OMNI.  The outward HP speed at this time is, in fact, close to the outward speed of V1.

283    **Summary and Conclusions**

284         We have extended our earlier calculations of the distance to the HTS which

285    covered the time period from the launch of V1 and V2 in 1977 to 2005 (Webber, 2005),

286    to the time period from 2006-2012.  We continue to use the V2 plasma data up to

287    2007.40 just before V2 reached the HTS.  In our updated calculations for the time period

288    between 2007.4-2011 we use the solar wind data from the SWEPAM instrument on the

289    ACE spacecraft and the OMNI data.

290         We find that to match the HTS crossing distances of 94.0 AU for V1 and 83.7 for

291    V2 we require a normalization factor of 0.92 between the N and S hemispheres.  This

292    factor represents a squashing of the heliosphere in the S hemisphere relative to the N as

293    originally suggested by Stone, et al., 2008 and Opher, et al., 2007.

294         We also find that, in response to a decreasing solar wind ram pressure, the HTS

295    moved inward after about mid 2008 reaching a minimum distance ~80 AU at the location

296    of V1 and 74 AU at V2 in early 2011.  Recall that the HTS crossed V1 at 94 AU in late

297    2004 and it crossed V2 at 84 AU at 2007.66, so this represents a shrinking of this

298    heliospheric parameter by ~15% at that time.  The HTS distance will start to increase

299    early in 2011 as a result of the increasing solar activity and solar wind pressure observed

300    by OMNI.

301         Our studies reveal that the HTS crossing of V2 at 2007.66 may not be unique.  It

302    is quite likely that structure just upstream of the HTS modified the solar wind ram

303    pressure.  As a result, instead of decreasing as previous calculations using the actually

304    measured solar wind ram pressure, the HTS distance at the V2 location remained almost

305    constant during the latter part of 2007.  The arrival of the IP shock from the December

306    2006 events on the Sun between about 2007.50 and 2007.84 thus pushed the HTS

307 outward just enough so that it almost reached V2 thus providing a possibility of a second
308 V2 crossing (or near crossing) after the initial crossing.

309 The distance to the HP is more uncertain than the HTS. Assuming that the ratio
310 of the HP distance to that of the HTS remains constant at a value between 1.4-1.6 during
311 the inward movement of the HTS, and allowing for a further 1 year delay for the HP to
312 reach an equilibrium distance relative to the HTS, if the HP/HTS ratio is only 1.4, V1
313 could encounter the HP as early as the beginning of 2011 at a distance of ~116 AU.

314 The HP continues to move inward until the early part of 2012 when it begins to
315 move outward due to the onset of the new solar cycle. At 2012.5 V1 will be at 121 AU.
316 This is equivalent to a distance ~1.50 times the corresponding HTS location one year
317 earlier.

318 So a non-encounter during the 2011-2012 time period would mean that the
319 equilibrium HP/HTS distance ratio is greater than 1.50.

320 At a nominal HP distance of 144 AU implied by a 1.6 HP/HTS distance ratio and
321 an HTS location of 90 AU, the HP would then not be encountered by V1 until 2019,
322 which is at the limit of the expected lifetime of V1.

323 So will V1 reach the HP and interstellar space during 2011-2012 or will this
324 illustrious mission be denied its final triumph? Or will it defy its makers once again and
325 last just long enough to cross the HP and live to send the data back to Earth?

326

**REFERENCES**

Burlaga, et al., (2008), Magnetic fields at the solar wind termination shock, Nature, <u>454</u>, 75-77

Cummings, A.C., et al., (2010),

Decker, R.B., S.M. Krimigis, E.C. Roelof and M.E. Hill, (2010), Variations of low energy ion distributions measured in the heliosheath, AIP Conf. Proc., <u>1352</u>, 51-57, Eds. Le Roux, Florinski, Zank and Coates

Decker, R.B., et al., (2010),

Ferreira, S.E.S, M.S. Potgieter and K. Scherer, (2007), Transport and acceleration of anomalous cosmic ray in the inner heliosheath, JGR, <u>112</u>, A11101, doi:10.1029/2007JA012477

Intriligator, D.S., W. Sun, M. Dryer, C.D. Fry, C. Deehr and J. Intriligator, (2005), From the Sun to the outer heliosphere: Modeling and analyses of the interplanetary propagation of the October/ November (Halloween) 2003 solar events, JGR, <u>110</u>, A09S10, doi:10.1029/ 2004JA010939

Intriligator, D.S., et al., (2010a), Voyager 2 high energy ions near the outward moving termination shock, AIP Conf. Proc., <u>1352</u>, 148-157, Eds. Le Roux, Florinski, Zank and Coates

Intriligator, D.S., et al., (2010b), Higher-energy plasma ions found near the termination shock: Analyses of Voyager 2 data in the heliosheath and in the outer heliosphere, JGR, <u>115</u>, A07107, doi:10.1029/2009JA014967

Izmodenov, V., G. Gloeckler and Y. Malaga, (2003), When will Voyager 1 and 2 cross the termination shock?, Geophys. Res. Lett., <u>30(7)</u>, 1351, doi:10.1029/2002GL016127

Florinski, V. and N.V. Pogorelev, (2009), Four dimensional transport of galactic cosmic rays in the outer heliosphere and heliosheath, Ap.J., <u>701</u>, 642-651

Lu, J.Y., Y.C. Whang and L.F. Burlaga, (1999), Interaction of a strong interplanetary shock with the termination shock, J. Geophys. Res., <u>104</u>, 28,249-28,254

McDonald, F.B., W.R. Webber and D.V. Reames, (2010), Unusual time histories of galactic and anomalous cosmic rays over the deep solar minimum of cycle 23/24, Geophys. Res. Lett., <u>37</u>, L18101, doi:10.1029/2010GL044218

357 Mueller, H.R., et al., (2006), Effects of variable interstellar environments on the heliosphere,

358     AIP Conf. Proc., 858, 33-38, Eds. Heerikhuisen, Florinski, Zank and Pogorelov

359 Mewaldt, R.A., A.J. Davis, K.A. Lave, R.A. Leske, E.C. Stone, M.E. Wiedenbeck, W.R.

360     Binns, E.R. Christian, A.C. Cummings, G.A. de Nolfo, M.H. Israel, A.W. Labrador and

361     T.T. von Rosenvinge, (2010), Record-setting cosmic-ray intensities in 2009 and 2010,

362     Astrophys. J. Lett., 723, L1-L6

363 Opher, M., E.C. Stone, P.C. Liewer and T. Gombosi, (2006), Global asymmetry of the

364     Heliosphere, AIP Conf. Proc., 858, 45-50, Eds. Heerikhuisen, Florinski, Zank and

365     Pogorelov

366 Opher, M., (2010), Magnetic fields in the local ISM and the local bubble, AAS Meeting #216,

367     #201.03

368 Richardson, J.D., C. Wang and M. Zhang, (2006a), Plasma in the outer heliosphere and the

369     Heliosheath, AIP Conf. Proc., 858, 110-115, Eds. Heerikhuisen, Florinski, Zank and

370     Pogorelov

371 Richardson, J.D., et al., (2006b), Source and consequences of a large shock near 79 AU,

372     Geophys. Res. Lett, 33 L23107, doi:10.1029/2006GL027983

373 Richardson, J.D. et al., (2008),Cool heliosheath plasma and deceleration of the upstream solar

374     wind at the termination shock, Nature, 454, 63-66

375 Stone, E.C., et al., (2008), An asymmetric solar wind termination shock, Nature, 454, 71-74

376 Wang, C. and J.W. Belcher (1999), The heliospheric boundary response to large-scale solar

377     wind fluctuations: A gas dynamic model with pickup ions, J. Geophys. Res., 104, 549-

378     556

379 Whang, Y.C. and L.F. Burlaga, (1993), Termination shock: solar cycle variations of location

380     and speed, J. Geophys. Res., 98, 15,221-15,230

381 Washimi, H., G.P. Zank, Q. Hu, T. Tanaka and K. Munakata, (2007), A forecast of the

382     heliospheric termination-shock position by three-dimensional MHD simulations, Ap.J.,

383     670, L139-L142

384 Washimi, H., G.P. Zank, Q. Hu and T. Tanaka, (2010), Realistic and time-varying outer

385     heliospheric modeling by three-dimensional MHD simulation, Ap.J., (submitted 2010)

386  Webber, W.R., (2005), An empirical estimate of the heliospheric termination shock location
387       with time with application to the intensity increases of MeV protons seen at Voyager 1 in
388       2002-2005, J. Geophys. Res., 110, A10103, doi:101029/2005JA011209
389  Webber, W.R., A.C. Cummings, F.B. McDonald, E.C. Stone, B. Heikkila and N. Lal, (2007),
390       Passage of a large interplanetary shock from the inner heliosphere to the heliospheric
391       termination shock and beyond: Its effects on cosmic rays at V1 and V2, Geophys. Res.
392       Letter, 34, , doi:10.1029/2007GL31339
393  Webber, W.R., A.C. Cummings, F.B. McDonald, E.C. Stone, B. Heikkila and N. Lal, (2009),
394       Transient intensity changes of cosmic rays beyond the heliospheric termination shock as
395       observed at Voyager 1, JGR, 114, A07108, doi:10.1029/2009JA014156
396
397

**FIGURE CAPTIONS**

398

399 **Figure 1:** Solar wind ram pressure measured at V2 (corrected to 1 AU) in 2005-2007. The
400 solar wind ram pressure profiles after 2007.4, a and b and the red line are discussed in the
401 text.

402 **Figure 2:** 27 day average solar wind ram pressure at the Earth from ACE-SWEPAM and
403 OMNI experiments, 1998-2010. Data smoothed with 5 period running average.

404 **Figure 3:** 26 day average solar wind ram pressure at V2 from 1998-2007 (corrected to 1
405 AU). V2 data is smoothed in the same way as Earth data. SWEPAM/OMNI data are
406 shown with an average time delay of 0.87 year.

407 **Figure 4:** Ratio of solar wind ram pressures measured by V2 and by SWEPAM from 1998 to
408 2007.4.

409 **Figure 5:** Calculated distance of the HTS at V1 for the time period 2002-2007, and V2 from
410 2005-2007. The HTS distance at V2 is normalized so that V2 crosses the HTS at
411 2007.66.

412 **Figure 6:** Calculated distance of the HTS at the location of V2 during 2007. The HTS
413 distance for the different solar wind ram pressure profiles a and b and the red line is
414 discussed in the text. The original V2 data profile is shown as a black line. The lines a
415 and b and the red line have been renormalized so that V2 crosses the HTS at 2007.66

416 **Figure 7:** Distance to the HTS and the HP at V1 for the time period 2007-2012. The
417 estimated location of the HP for this same time period is shown by the shaded region. It
418 is assumed that the ratio of the HP distance to the HTS distance, delayed by 1 year,
419 remains a constant with time with values between 1.4-1.6 during this time period.

420

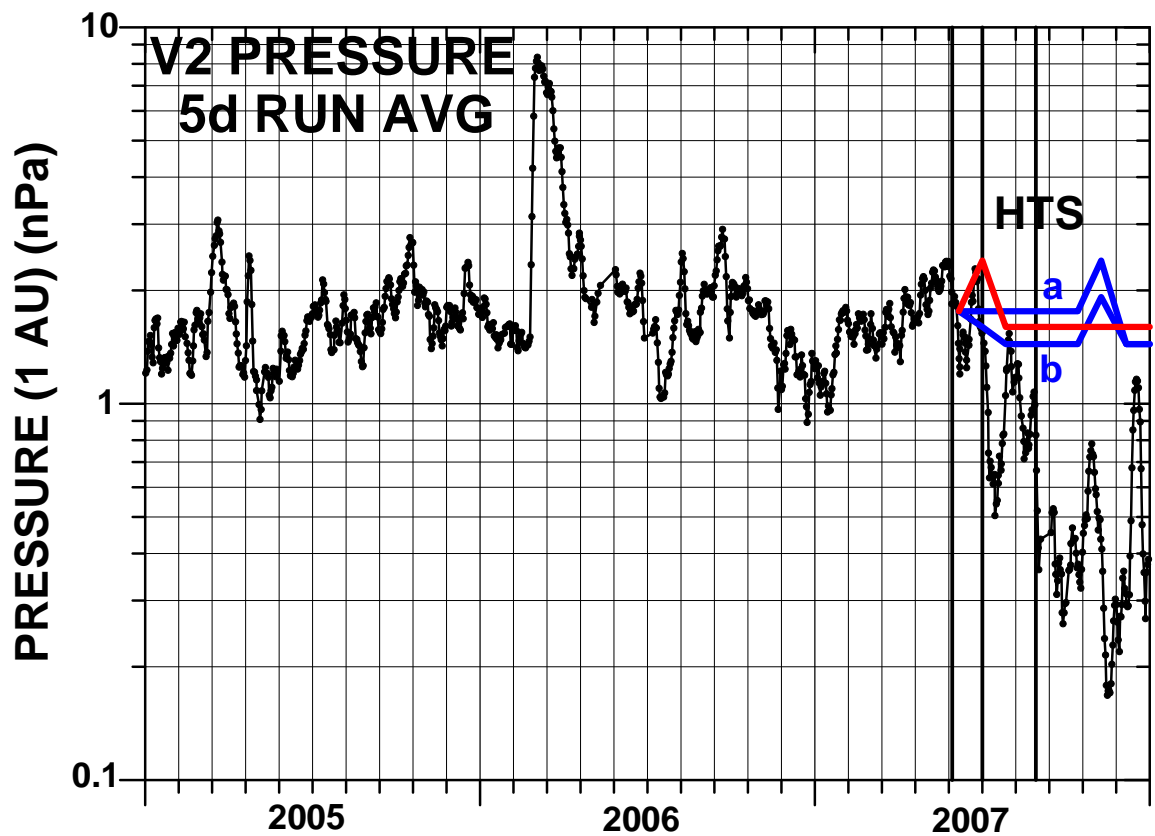

FIGURE 1

421

422

423

424

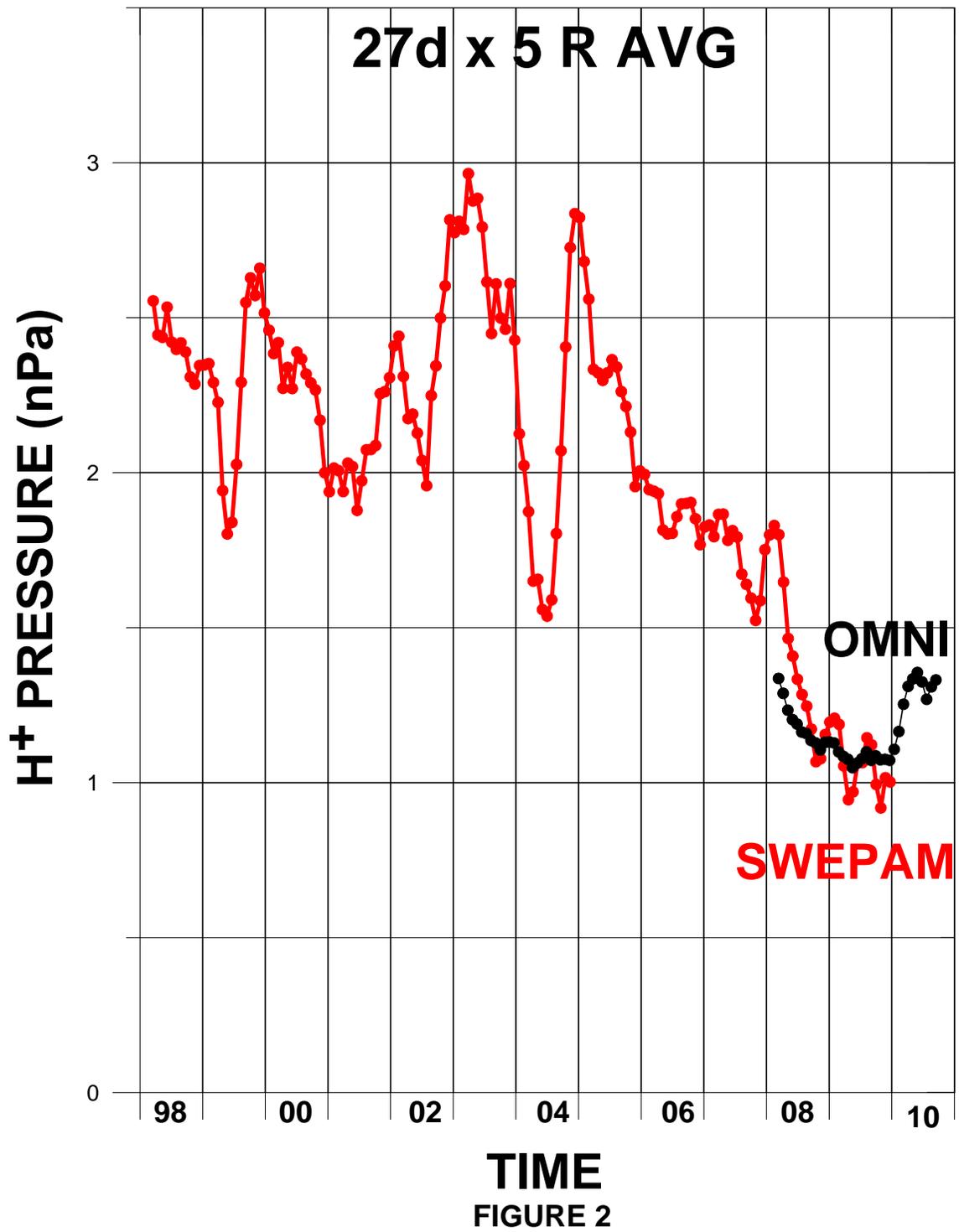

FIGURE 2

425

426

427

428

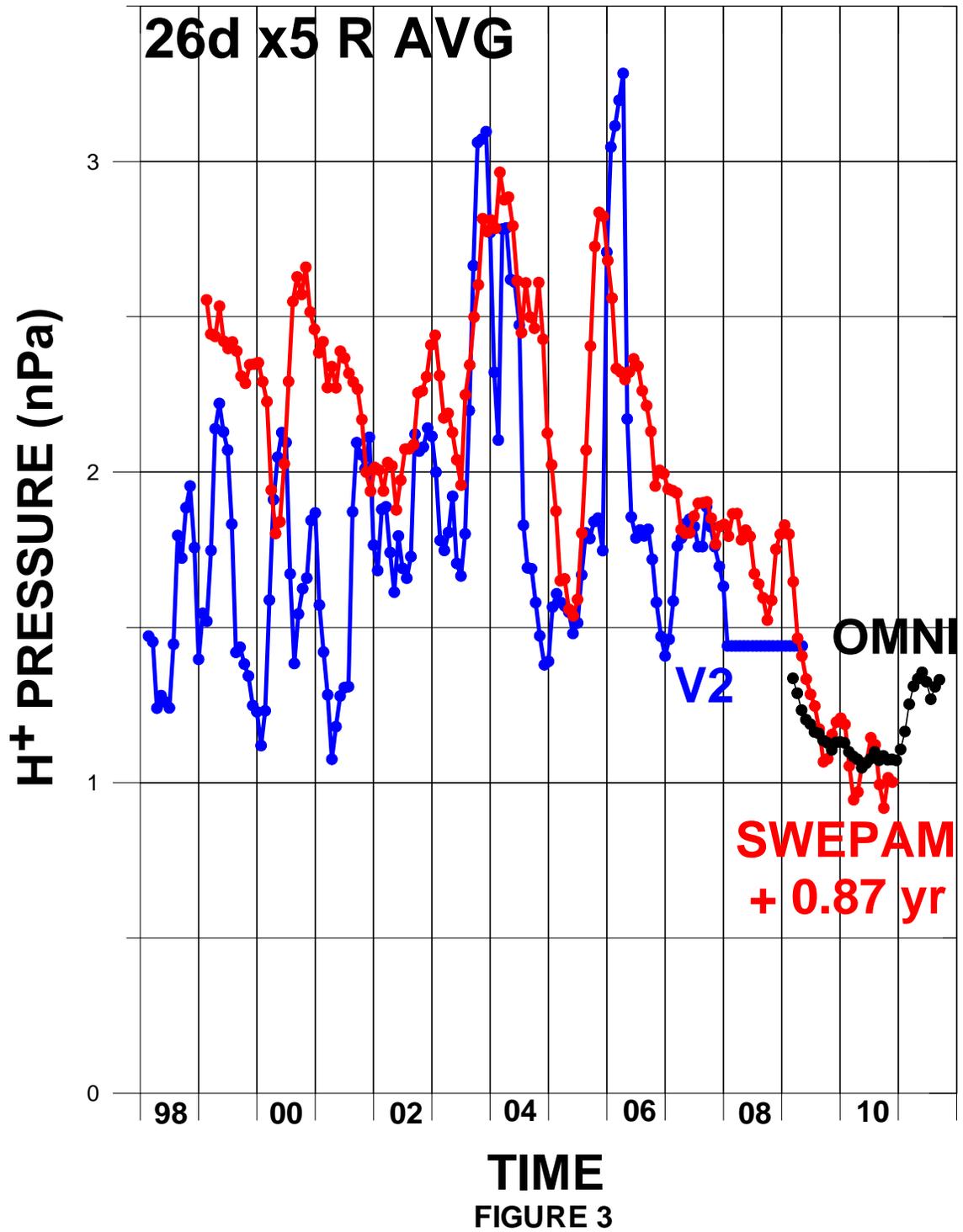

**TIME**

FIGURE 3

429

430

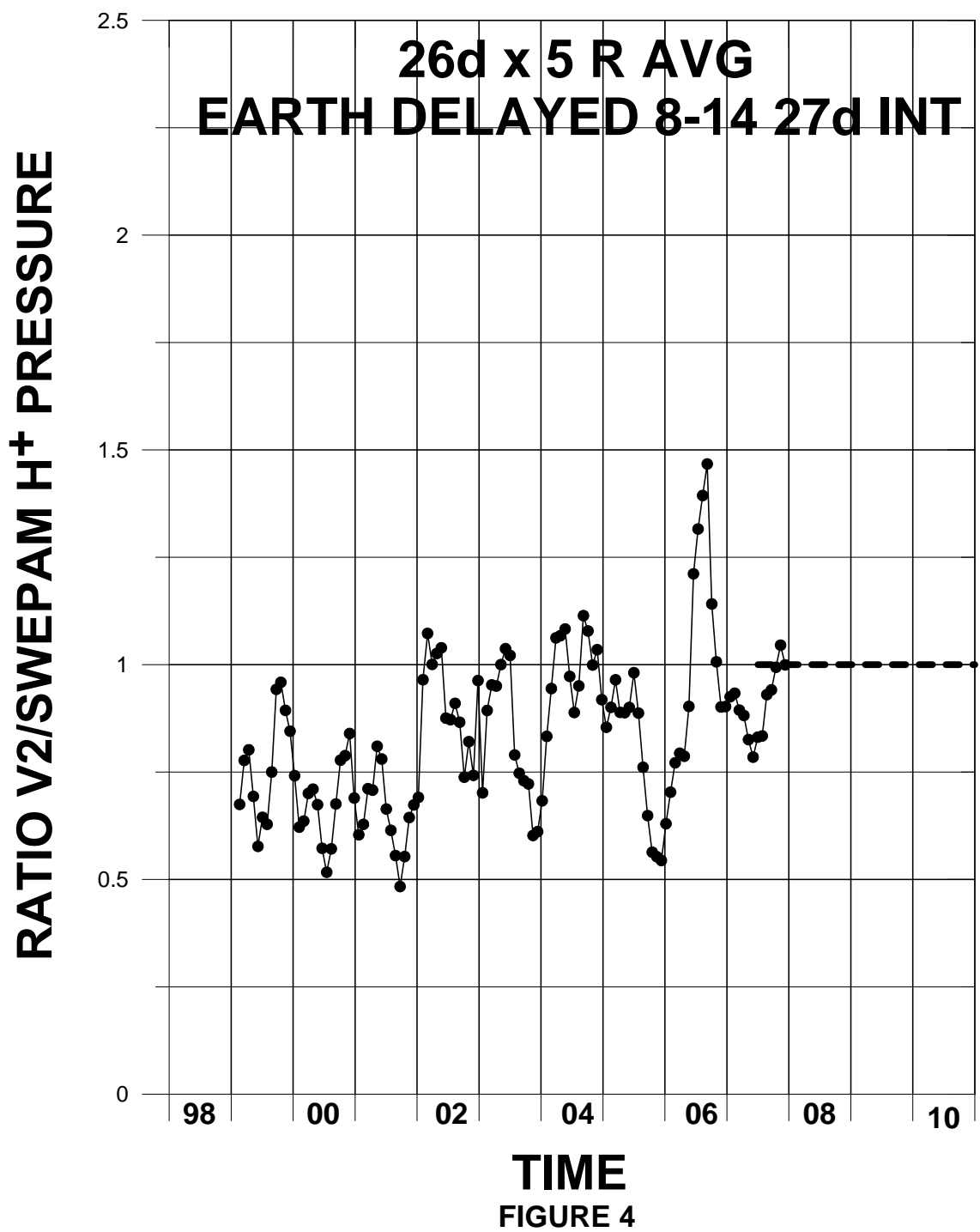

FIGURE 4

431

432

433

434

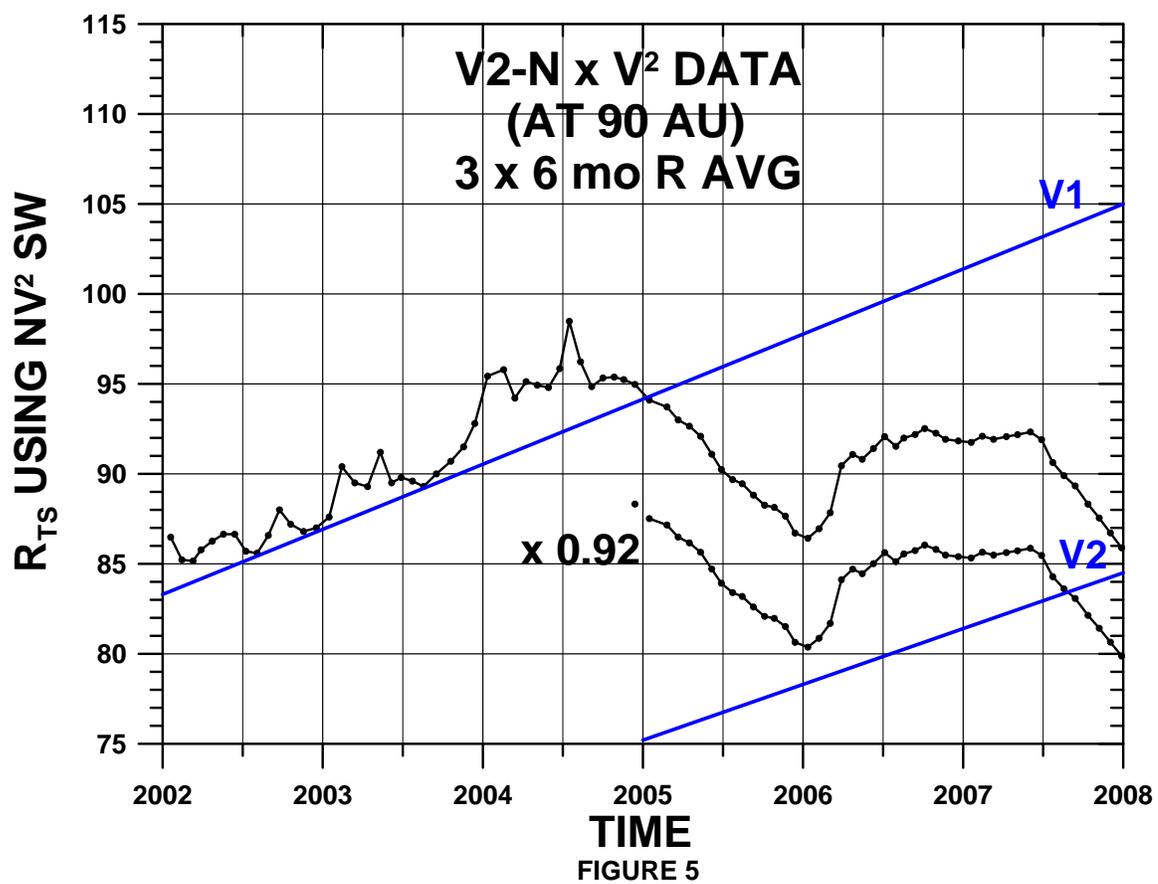

435

436

437

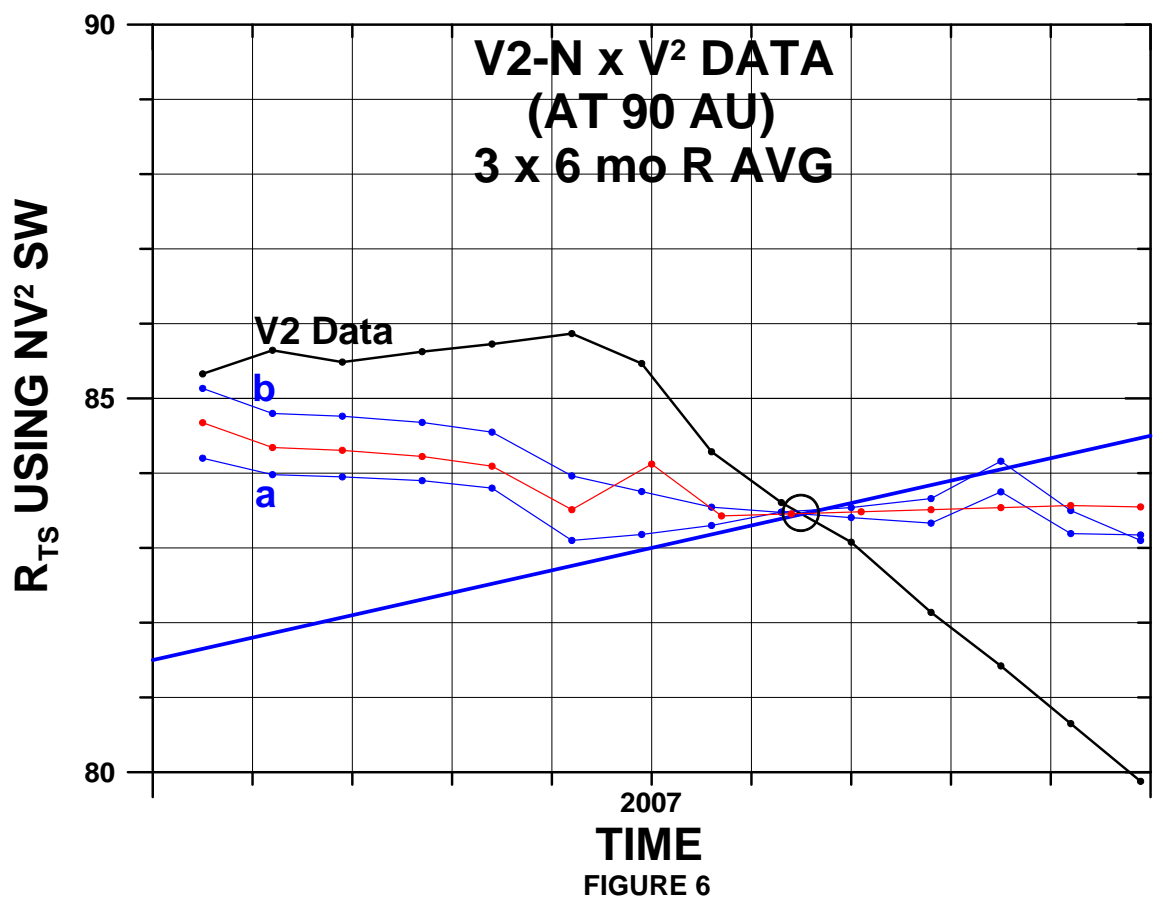

FIGURE 6

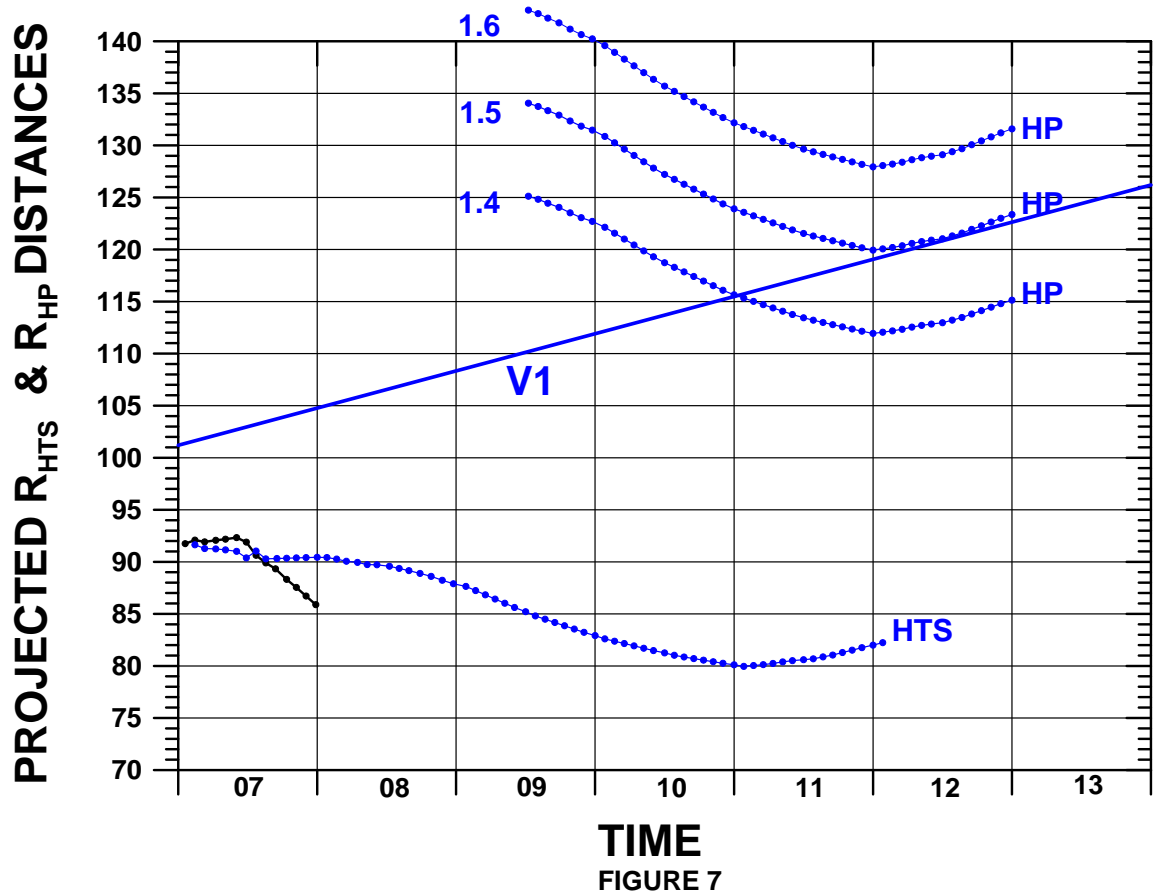

**TIME**

**FIGURE 7**